\documentclass[11pt]{article} 
\usepackage{graphicx}
\usepackage{hyperref}
\usepackage{amssymb}
\usepackage{authblk}

\date{\today}

\title{\bf A search for Secluded Dark Matter in the Sun with the ANTARES neutrino telescope}

\author[a]{S.~Adri\'an-Mart\'inez,\footnote{Corresponding author, e-mail: siladmar@upv.es}}
\author[b]{A.~Albert}
\author[c]{M.~Andr\'e}
\author[d]{G.~Anton}
\author[a]{M.~Ardid,\footnote{Corresponding author, e-mail: mardid@fis.upv.es}}
\author[e]{J.-J.~Aubert}
\author[f]{T.~Avgitas}
\author[f]{B.~Baret}
\author[g]{J.~Barrios-Mart\'{\i}}
\author[h]{S.~Basa}
% ...

\affil[a]{\scriptsize{Institut d'Investigaci\'o per a la Gesti\'o Integrada de les Zones Costaneres (IGIC) - Universitat Polit\`ecnica de Val\`encia. C/  Paranimf 1 , 46730 Gandia, Spain.}}
\affil[b]{\scriptsize{GRPHE - Universit\'e de Haute Alsace - Institut universitaire de technologie de Colmar, 34 rue du Grillenbreit BP 50568 - 68008 Colmar, France}}
\affil[c]{\scriptsize{Technical University of Catalonia, Laboratory of Applied Bioacoustics, Rambla Exposici\'o,08800 Vilanova i la Geltr\'u,Barcelona, Spain}}
\affil[d]{\scriptsize{Friedrich-Alexander-Universit\"at Erlangen-N\"urnberg, Erlangen Centre for Astroparticle Physics, Erwin-Rommel-Str. 1, 91058 Erlangen, Germany}}
\affil[e]{\scriptsize{Aix-Marseille Universit\'e, CNRS/IN2P3, CPPM UMR 7346, 13288 Marseille, France}}
\affil[f]{\scriptsize{APC, Universit\'e Paris Diderot, CNRS/IN2P3, CEA/IRFU, Observatoire de Paris, Sorbonne Paris Cit\'e, 75205 Paris, France}}
\affil[g]{\scriptsize{IFIC - Instituto de F\'isica Corpuscular , CSIC - Universitat de Val\`encia. C/ Catedr\'atico Jos\'e Beltr\'an 2, E-46980 Paterna, Valencia, Spain}}
\affil[h]{\scriptsize{LAM - Laboratoire d'Astrophysique de Marseille, P\^ole de l'\'Etoile Site de Ch\^ateau-Gombert, rue Fr\'ed\'eric Joliot-Curie 38,  13388 Marseille Cedex 13, France}}
% ...

\begin{document} 


\maketitle 

\begin{abstract}
A search for Secluded Dark Matter annihilation in the Sun using 2007-2012 data of the ANTARES neutrino telescope is presented. Three different cases are considered: a) detection of dimuons that result from the decay of the mediator, or neutrino detection from: b) mediator that decays into a dimuon and, in turn, into neutrinos, and c) mediator that decays directly into neutrinos. As no significant excess over background is observed, constraints are derived on the dark matter mass and the lifetime of the mediator.
\end{abstract}
%


\section{Introduction}
\label{sec:intro}

There is strong cosmological and astrophysical evidence for the existence of Dark Matter (DM) in the Universe. The observations indicate that DM, about 26\% of the total mass-energy of the Universe, is non-baryonic, non-relativistic and not subject to electromagnetic interactions \cite{01}. In the framework of the Weakly Interacting Massive Particles (WIMPs) paradigm, the visible baryonic part of a galaxy is embedded in the DM halo. In the most common scenario, WIMPs can scatter elastically with matter and become trapped in massive astrophysical objects such as the Sun\cite{02}. There, DM particles could self-annihilate, reaching equilibrium between capture and annihilation rates over the age of the Solar System. The standard scenario assumes that the products of DM annihilation are Standard Model (SM) particles, which interact with the interior of the Sun and are largely absorbed \cite{03}. However, during this process, high-energy neutrinos may be produced, which can escape and be observed by neutrino detectors, such as ANTARES. Limits on WIMP DM annihilation in the Sun have been reported already in ANTARES \cite{1}, and in other neutrino telescopes: Baksan \cite{2}, Super-Kamiokande \cite{3} and IceCube \cite{4,4b}. 

An alternative hypothesis is based on the idea that DM is ``secluded'' from SM particles and that the annihilation is only possible through a metastable mediator ($\phi$), which subsequently decays into SM states \cite{5,6,7,8,9}. These models retain the thermal relic WIMP DM scenario while at the same time  explain the positron-electron ratio observed by PAMELA \cite{10}, FERMI \cite{11}, and measured recently by AMS-II with improved accuracy \cite{12,12b}. In the Secluded Dark Matter (SDM) scenario, the presence of a mediator dramatically changes the annihilation signature of DM captured in the Sun. If the mediators live long enough to escape the Sun before decaying, they can produce fluxes of charged particles, $\gamma$-rays or neutrinos \cite{13,14} that could reach the Earth and be detected. In many of the secluded dark matter models, $\phi$ can decay into leptons near the Earth. The signature of leptons arising from $\phi$ decays may differ substantially from other DM models. Assuming that the DM mass ($\sim$ 1 TeV) is much greater than the $\phi$ mass ($\sim$ 1 GeV) the leptons are boosted. If these leptons are muons, which is the less constrained case to explain the positron-electron ratio \cite{15,16,17}, the signature in the vicinity of the detector would be two almost parallel muon tracks. In Ref. \cite{18} this possibility is discussed and the expected sensitivity for the IceCube neutrino telescope is calculated. Why a neutrino telescope generally interprets the dimuon signature as a single muon, the different energy deposition can help to discriminate this case from a muon induced by a neutrino interaction \cite{19}. On the other hand, even for short-lived mediators that decay before reaching the Earth, neutrinos from the products of mediator decays could be detected in neutrino telescopes as well. Finally, another possibility is that mediators may decay directly into neutrinos \cite{20}. In this case, the neutrino signal could be enhanced significantly compared to the standard scenario even for quite short-lived mediators. The mediators will be able to escape the dense core of the Sun where high-energy neutrinos can interact with nuclei and be absorbed. The fact that the solar density decreases exponentially with radius facilitates neutrinos injected by mediators at larger radii to propagate out of the Sun. 

In this work an indirect search for SDM using the 2007-2012 data recorded by the ANTARES neutrino telescope is reported. The analysis treats the different mediator decay products: 
\begin{itemize}
 \item[a)] direct detection of dimuons 
 \item[b)] neutrinos from decays of dimuons produced by mediators that decay before reaching the Earth 
 \item[c)] neutrinos produced by mediators that decay directly to neutrinos and antineutrinos. 
\end{itemize}
The analysis procedure is similar to the previous ANTARES search for DM annihilation in the Sun \cite{1}, but optimising the search for the expected signal in the case of SDM. In the following, neutrino will mean neutrino plus anti-neutrino, unless explicitly stated otherwise.

\section{The ANTARES neutrino telescope}\label{sec:antares}

The ANTARES neutrino telescope \cite{21} is located in the Mediterranean Sea, at a depth of 2.5 km, about 40 km offshore from Toulon (France). It is presently the largest neutrino telescope in the Northern hemisphere and consists of 885 Optical Modules (OMs) arranged in a three-dimensional array. The operation principle is based on the detection of the Cherenkov light induced by relativistic charged particles produced in interactions of high-energy neutrinos in the surroundings of the detector. The OMs are installed along 12 lines anchored to the sea floor and kept vertical by a submerged buoy. The length of the lines is 450 m and the distance between the lines is 60--75 m. The OMs are grouped in triplets in order to reduce the effect of optical background produced by Potassium-40 decays and bioluminescence. A line comprises 25 triplets separated by a vertical distance of 14.5 m. The position \cite{22,22b} and time \cite{23} information of the photons detected by the OMs are used to determine the muon direction. The reconstruction algorithm is based on the minimisation of a $\chi^2 -$like quality parameter of the track reconstruction, Q, which uses the difference between the expected and measured times of the detected photons, taking into account the effect of light absorption and scattering in the water \cite{24}. Fig. 3 of Ref. \cite{1} shows the distribution of reconstruted events as a function of Q.

The installation of the detector was completed in 2008, although during 2007 five lines were already installed and data taking begun. In this analysis data recorded between the $27^{\rm th}$ of January 2007 and the $31^{\rm st}$ of October 2012 are used, corresponding to a total livetime of 1321 days, without taking into account the visibility of the Sun. During this time, the detector consisted of 5 lines for most of 2007 and of successively 10 and 12 lines from 2008 to 2012.

\section{Signal and background estimation}\label{sec:signal}

Dimuon and neutrino-induced candidates are events reconstructed as upgoing. The duty cycle of this search for events from the Sun corresponds to 50\% of the detector livetime. Two main sources of background are present in the ANTARES data: 1) Downgoing atmospheric muons resulting from the interaction of cosmic rays in the atmosphere. These background events are strongly reduced by the deep sea location and by the reconstruction algorithms that are optimised for upgoing events. Cuts on the quality of the tracks are also applied to reject downgoing muons mis-reconstructed as upgoing. 2) Atmospheric neutrinos produced by cosmic rays. These neutrinos can traverse the Earth, so they can be detected as upgoing tracks and cannot be rejected on an event-by-event basis. Both kinds of background have been simulated and good agreement with data has been found \cite{1}. Nevertheless, the background estimation is performed using scrambled data, by randomising the time of selected events, to reduce the effect of systematic uncertainties (efficiency of the detector, assumed atmospheric fluxes, etc.). 

To evaluate the sensitivity of ANTARES to the signal from SDM models for the case a) in which dimuons are detected directly, a new tool for dimuon signal generation (DiMugen) has been developed \cite{25}. DiMugen generates and propagates dimuons produced by the decay of mediators resulting from DM annihilation. For this analysis, the mediator arrives from the Sun's position during the period under study. Different DM masses in the range between 30 GeV to 10 TeV have been simulated, using in most cases a typical mass of 1 GeV for the mediator $\phi$. Once the muons are generated in the vicinity of the detector, simulations of the travel and interactions of muons are performed, as well as the detection of the Cherenkov light by the optical modules. Triggering and reconstruction algorithms are also included in the process in order to evaluate the global efficiency for the detection of dimuons as a function of the quality parameter, Q, and the half-cone angle from the Sun direction, $\Psi$.
      
To evaluate the ANTARES sensitivity for the cases where the neutrino is the final decay product that arrives at Earth, the ANTARES effective areas for neutrinos as a function of the Q and $\Psi$ according to neutrino simulations have been used. For this, it is necessary to know the energy spectra of neutrinos arriving at the detector. In case b) the neutrino spectra have been obtained from Michel's spectra of neutrinos from muon decay and taking into account the boost \cite{25}. For scenario c) and assuming long mediator lifetimes with respect to the time required to exit from the Sun's core, the neutrino spectra are almost flat in the energy range under study \cite{20}. For these cases the assumption that after oscillations all neutrino flavours arrive at Earth with the same proportion has been made. For case b), this results in a factor of 2/3 for the muon neutrinos with respect to the parent muons, since a muon neutrino and an electron neutrino result from a muon decay, and these neutrinos are spread to all flavours. There is not any change in the neutrino flavour composition of case c) due to oscillations since it is assumed that all kinds of neutrinos result from the mediator decays.

\section{Optimisation of the event selection criteria}\label{sec:optimi}

In order to avoid any bias in the event selection, a blinding policy has been followed using as observables the angular separation of the track with respect to the Sun's direction and the track quality parameter $Q$. The values of the cuts have been chosen before looking at the region where the signal is expected. The best sensitivities for dimuon (or neutrino) fluxes and cross sections are extracted using the Model Rejection Factor (MRF) method \cite{26}. It consists of finding the set of cuts which provide, on average, the best flux upper limit taking into account the observed background and the efficiency for a possible signal flux. The procedure to obtain the sensitivity for a neutrino flux coming from DM annihilations in the Sun is described in Ref. \cite{1}. The detection of a couple of muons arising from the decay of the mediators is performed using the same reconstruction technique described in Ref. \cite{1} and used for single muons. A couple of parallel muons with relative separation smaller than the distance of strings in the detector yields a number of Cherenkov photons practically indistinguishable from that of a single muon. For this reason, the case a) is treated in the same way as cases b) and c).  

\begin{figure}[tbp]
\begin{center}
\includegraphics[width=12cm]{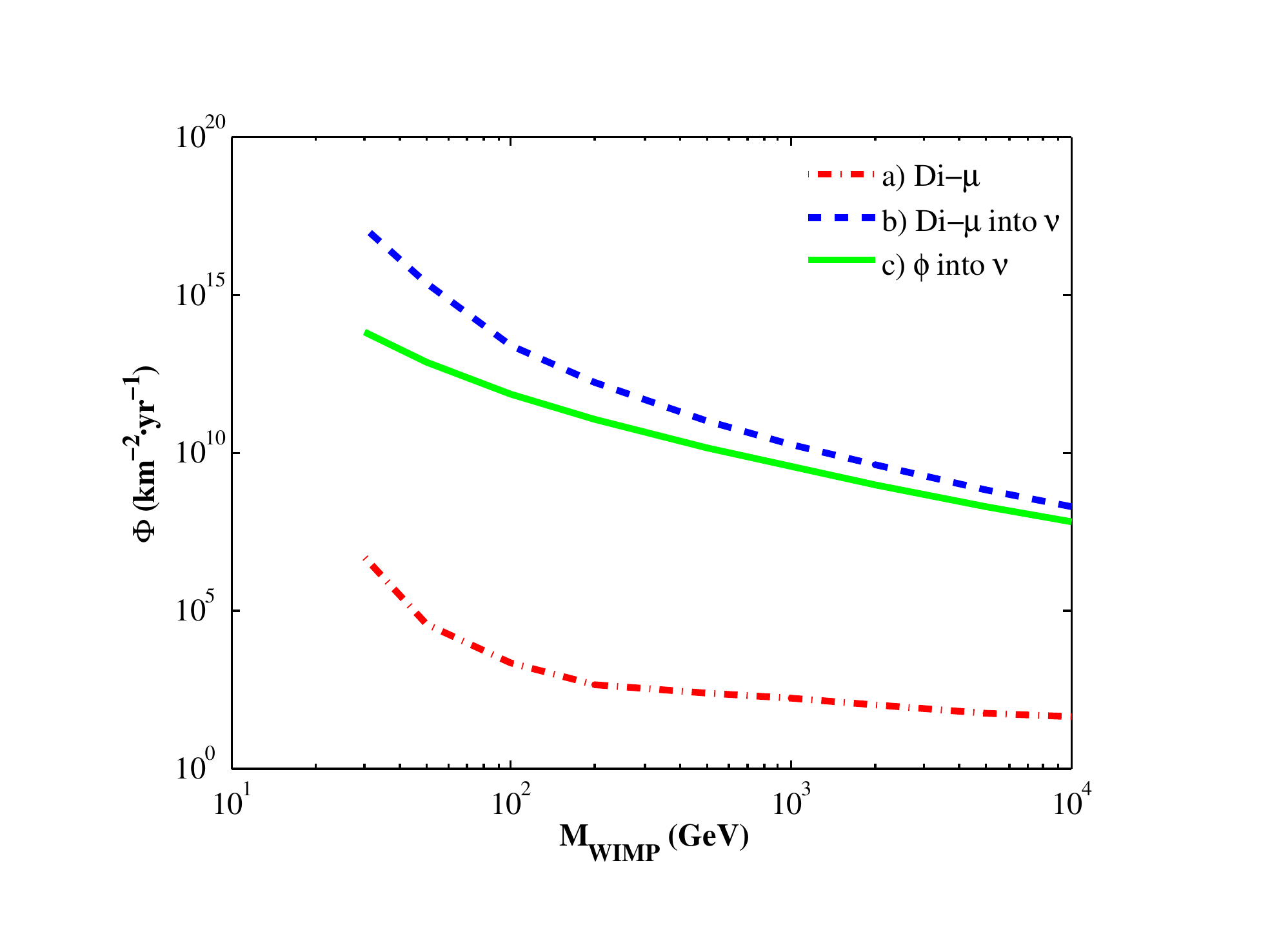}
\end{center}
\caption{\label{fig:1} Sensitivity, expected 90\% confidence limit, for DM annihilation in the Sun for three SDM scenarios with the data recorded by ANTARES between 2007 and 2012.}
\end{figure}

Then, the MRF is used to determine the best value for the cut on the half-cone angle around the Sun ($\Psi_{\rm cut}$) and the track quality cut parameters ($Q_{\rm cut}$) for the different cases (a, b, c) and the different DM masses studied. Since in most cases the difference in flux sensitivities between different optimisations is not large, it was decided to limit the optimisations to four different sets of cuts, see Table \ref{tab:1}, that were representative of all possible situations, within a few percent sensitivity difference from the optimal one \cite{25}. There are three optimisation sets corresponding roughly to lower, intermediate and larger DM masses for the dimuon detection case. For the neutrino detection cases, the latter set is also used for larger DM masses and another additional optimisation is used for lower and intermediate DM masses. Sensitivities to the 
particle flux at Earth,$\Phi$, obtained for the different cases studied are shown in Figure \ref{fig:1}.

\section{Results and discussion} \label{sec:disc}

\begin{figure}[tbp]
\begin{center}
\includegraphics[width=10cm]{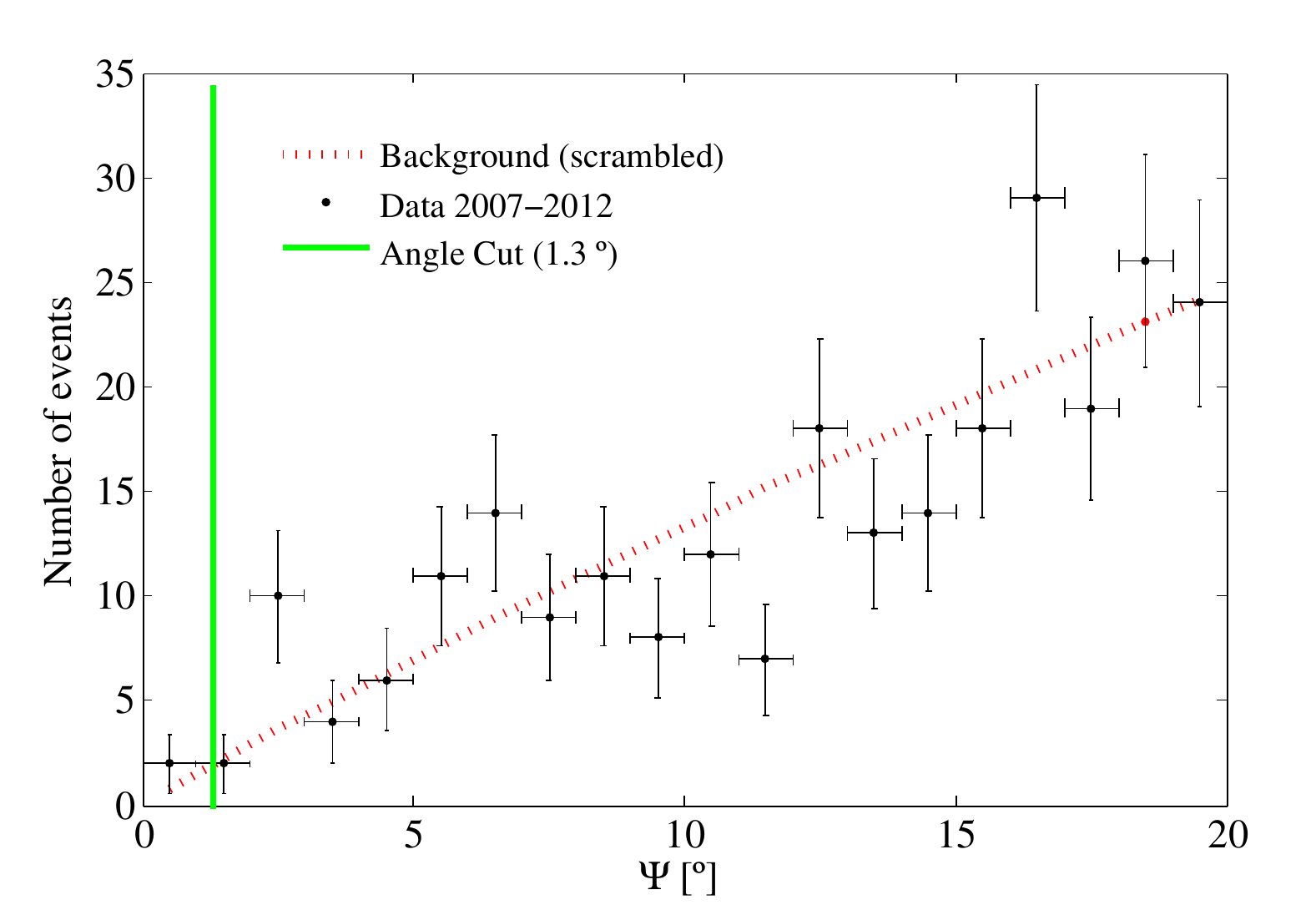}
\end{center}
\caption{\label{fig:2} Differential distribution of the angular separation of the event tracks with respect to the Sun's direction with $Q_{\rm cut}= 1.8$ for data (black) and expected background (red curve). The statistical uncertainties are shown for the data.}
\end{figure}

After the optimisation of the flux sensitivities using the MRF with scrambled data, the data coming from the Sun direction was unblinded. As an example, Figure \ref{fig:2} shows the distribution of events detected for $Q<1.8$ as a function of the angular distance from the Sun. Good agreement between data and the expected background obtained from scrambled data is observed. The green line indicates the angular cut selected for this analysis. Table 1 summarises the number of events observed and expected from background for the different sets of cuts. Since no significant excess is observed, the 90\% Confidence Level (CL) upper limit values in the Feldman-Cousins approach \cite{27} are used to constrain the model. 

\begin{table}[tbp]
\centering
\begin{tabular}{ccccccc}
\hline
Q$_{\rm cut}$ & $\Psi_{\rm cut}$ & SDM scenario & DM mass & N$_{\rm obs}$ & N$_{\rm back}$ & $\mu_{90\%}$ Upper Limit\\
\hline
1.8 & 1.3 & a), b), c) & Large & 2 & 1.25 & 4.7\\
1.6 & 1.4 & a) & Intermediate & 2 & 0.89 & 5.0\\
1.6 & 2.0 & a) & Low & 3 & 1.8 & 5.6\\
1.4 & 2.6 & b), c) & Intermediate, Low & 3 & 2.0 & 5.4 \\
\hline
\end{tabular}
\caption{\label{tab:1} Number of observed events, N$_{\rm obs}$, and expected background, N$_{\rm back}$, for the sets of cuts (track quality parameter, Q$_{\rm cut}$, and half-cone angle around the Sun, $\Psi_{\rm cut}$) used for the different SDM scenarios and DM mass. The resulting 90\% confidence level upper limit is also shown ($\mu_{90\%}$).}
\end{table}

Taking into account the upper limits obtained in Table \ref{tab:1} and the corresponding effective areas, the upper limits on the dimuon and neutrino fluxes at Earth, $\Phi$, are derived, which almost coincide with the sensitivities shown in Figure \ref{fig:1}. Following the arguments given in Ref. \cite{18}, the dimuon (or neutrino) flux at Earth can be translated into DM annihilation rate in the Sun for scenarios a) and b). Assuming a 100\% branching ratio for the $\phi\rightarrow\mu^+ +\mu^-$ decay channel of the mediator, and taking into account the solid angle factor and the decay probabilities, the following relationships between the annihilation rate, $\Gamma$, and the dimuon and muonic neutrino fluxes, respectively $\Phi_{\mu\mu}$ and $\Phi_\nu$ \cite{25}, are obtained: 
\begin{itemize}
\item[Case a)] 
\begin{equation}\label{gamma1}
\Gamma=\frac{4\pi D^2\Phi_{\mu\mu}}{2e^{-D/L}(1-e^{-d/L})},
\end{equation}
\item[Case b)] 
\begin{equation}\label{gamma2}
\Gamma=\frac{4\pi D^2\Phi_{\nu}}{\frac{8}{3}(e^{-R_{Sun}/L}-e^{-D/L})},
\end{equation}
\end{itemize}
where $d$ is a characteristic distance related to the detector size, $D$ is the distance between the Sun and the Earth; $R_{Sun}$ is the radius of the Sun; and $L$ is the mediator's decay length, $L=\gamma c \tau$, i.e. the product of the mediator's lifetime, $\tau$, the speed of light, $c$, and the relativistic boost factor $\gamma= m_{DM}/m_\phi$.

For the case in which mediators decay directly into neutrinos, only the situation in which the mediator lifetime is long enough has been considered, so that the absorption of neutrinos in the Sun becomes negligible. If the lifetime of the mediator is small, the final energy spectrum of neutrinos would be quite similar to the case of typical DM searches \cite{20}. For long-lived mediators, ($L>10^{5}$ km), in this scenario, the relationship between $\Gamma$ and $\Phi_\nu$ is \cite{25}:
\begin{itemize}
\item[Case c)] 
\begin{equation}\label{gamma3}
\Gamma=\frac{4\pi D^2\Phi_{\nu}}{\frac{4}{3}(1-e^{-D/L})}.
\end{equation}
\end{itemize}

\begin{figure}[tbp]
\centering  
\includegraphics[width=7.6cm]{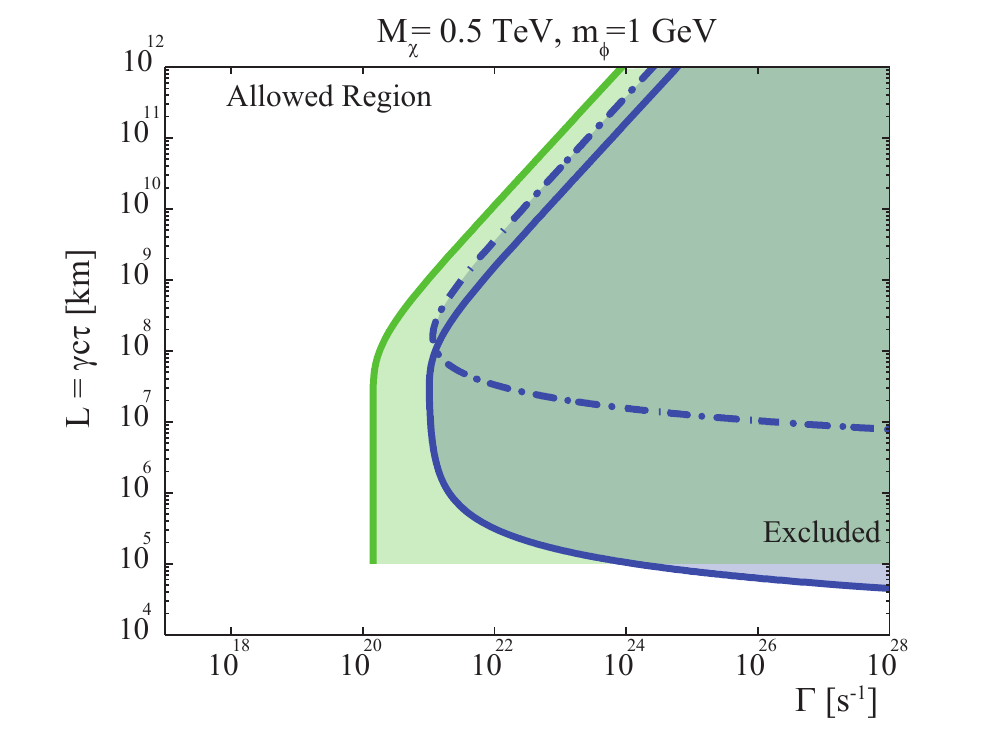}
\hfill
\includegraphics[width=7.6cm]{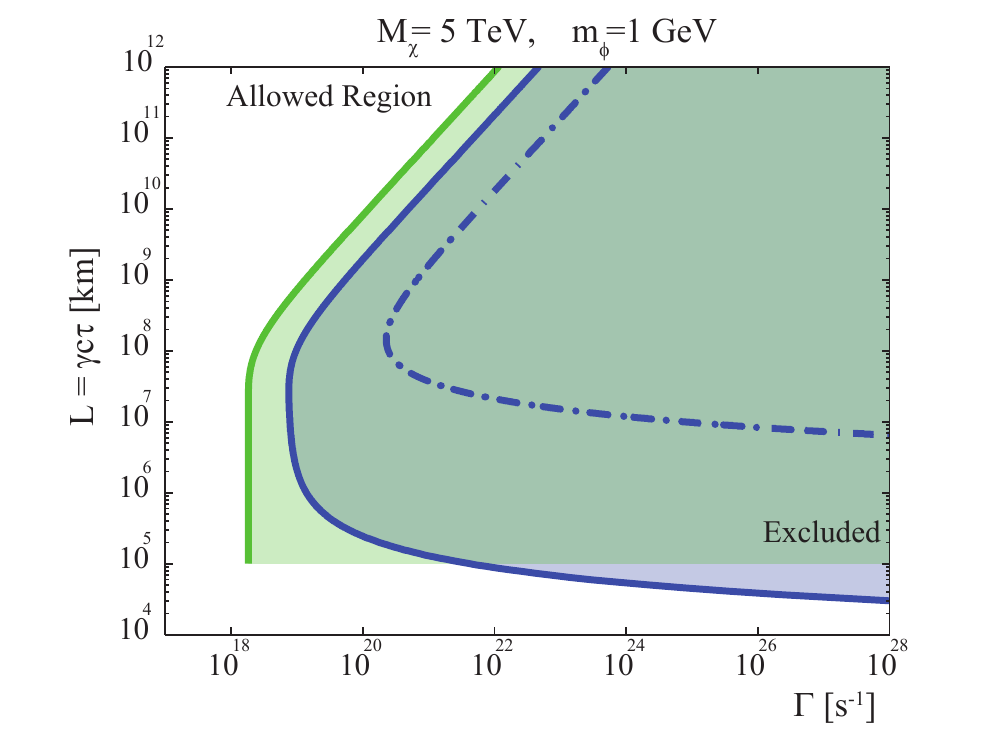}
\caption{\label{fig:3} ANTARES exclusion limits for the SDM cases studied by products of DM annihilation in the Sun through mediators decaying into: dimuons (dashed-dotted blue), neutrinos from dimuons (solid blue), directly into neutrinos (green) as a function of the annihilation rate ($\Gamma$) and the decay length ($L = \gamma c\tau$) for 0.5 and 5 TeV DM masses. The shadowed regions are excluded for these models.}
\end{figure}

Constraints on the annihilation rates as a function of mediator lifetime and dark matter mass have been obtained. For example, Figure \ref{fig:3} shows the ANTARES exclusion limits for the SDM scenarios for DM masses of 0.5 (left) and 5 (right) TeV using the typical $\phi$ mass of 1~GeV. Blue lines indicate the exclusion region in the di-muon case, either by direct detection (dotted-dashed line) or through detection of neutrinos (solid line). For large decay lengths $L$ ($L>D$), i.e. long mediator lifetime, the direct detection of dimuons is more efficient than neutrino detection for small DM masses, whereas the opposite holds for larger masses. The transition is around 0.8 TeV in DM mass. Naturally, for small $L$ ($L \ll D$) neutrino detection is much more efficient. Green lines indicate the exclusion regions of the parameter space for the scenario of SDM with mediator decaying into neutrinos. More stringent constraints are obtained in this scenario mainly due to the harder neutrino energy spectrum. 

\begin{figure}[tbp]
\centering  
\includegraphics[width=7.6cm]{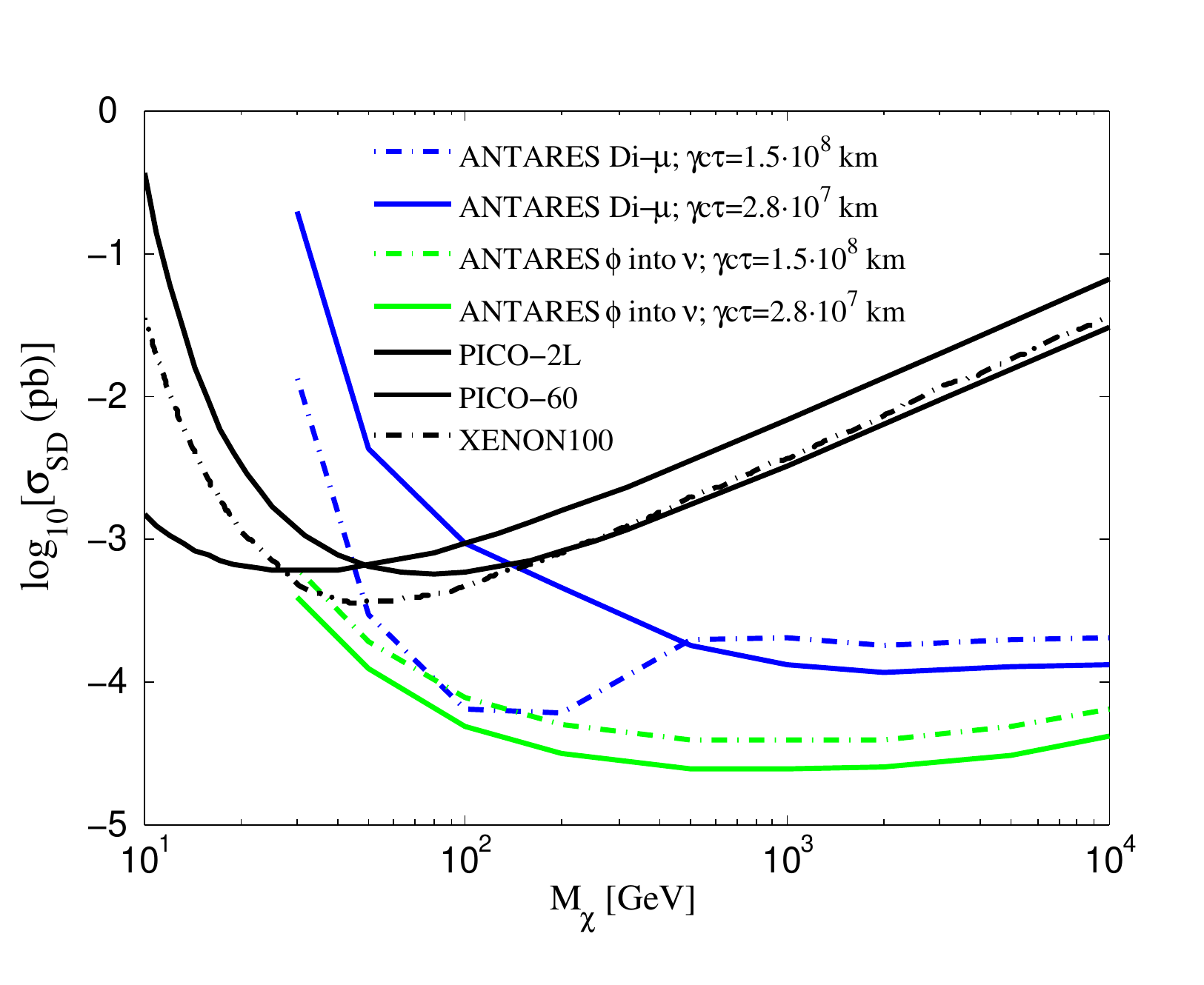}
\hfill
\includegraphics[width=7.6cm]{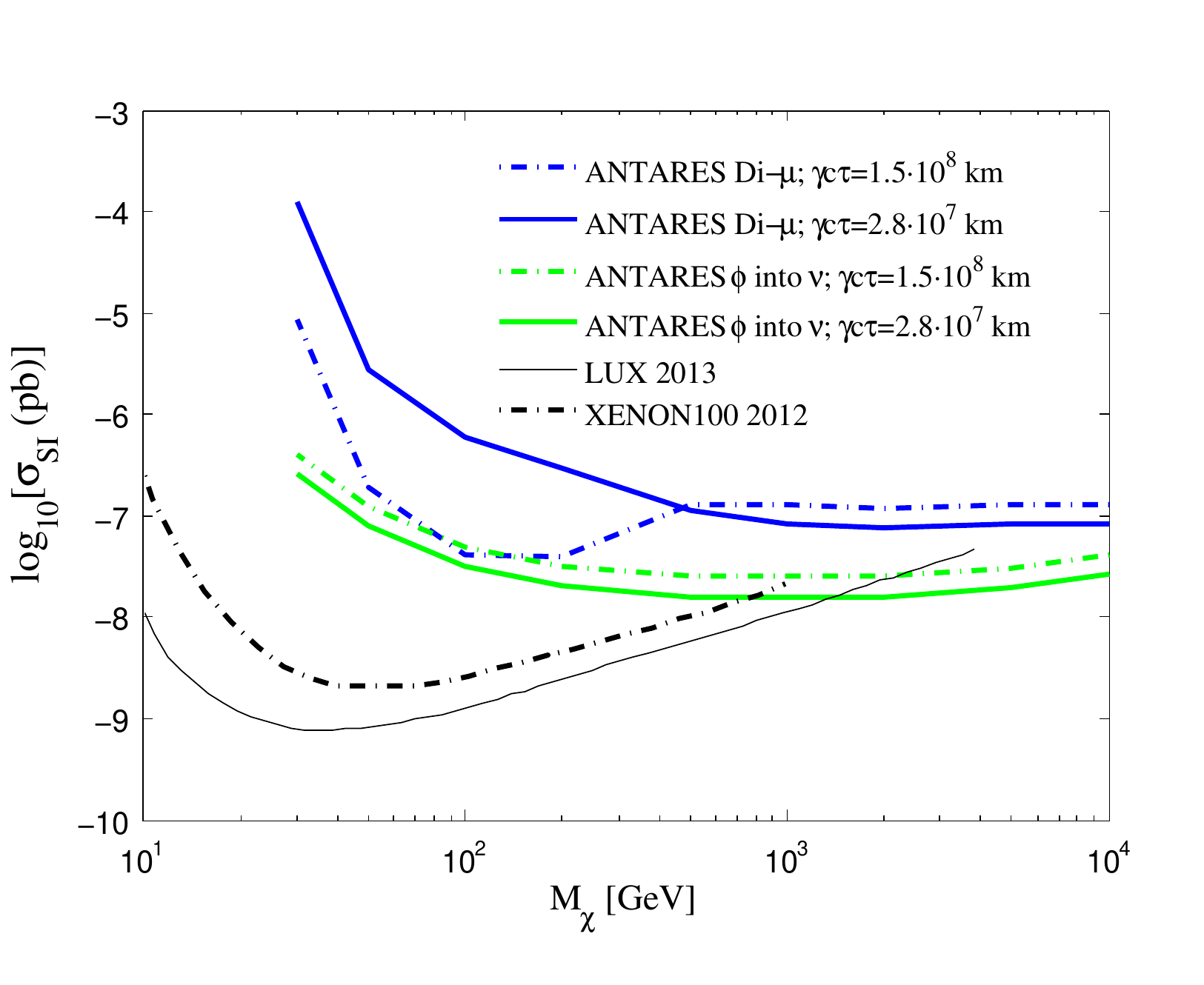}
\caption{\label{fig:4} ANTARES 90\% CL upper limits on WIMP-nucleon cross section as a function of WIMP mass. The left panel refers to spin-dependent and the right one to spin-independent WIMP interactions. Two favourable mediator lifetimes are considered. The current bounds from PICO \cite{30,30b,30c}, LUX \cite{31} and XENON \cite{32,33} are also shown.}
\end{figure}

Limits on the DM-nucleon interaction can also be derived for these cases. Assuming that, as expected, there is equilibrium of the DM population in the Sun, i.e., the annihilation balances the DM capture, $\Gamma =C_{DM}/2$ , and that according to \cite{28} the capture is approximately
\begin{equation}\label{cdm}
C_{DM}=10^{20} {\rm s}^{-1} \left(\frac{1 {\rm TeV}}{M_\chi}\right)^2  \frac{2.77\sigma_{SD}+4270\sigma_{SI}}{10^{-40} {\rm cm}^2} , 
\end{equation}
where, $\sigma_{SD}$ and $\sigma_{SI}$ are the spin-dependent (SD) and spin-independent (SI) cross sections, respectively, and $M_\chi$ is the DM mass. The limits on the SD and SI WIMP-proton scattering cross-sections are derived for the case in which one of them is dominant. The sensitivity in terms of annihilation rates depends on the lifetime of the mediator $\phi$. To assess the potential to constrain these models, lifetime values for which the sensitivities are optimal have been assumed. For the dimuon case, the lifetime has to be long enough to assure that the mediator reaches the vicinity of the Earth, so mediators with a decay length of about the Sun-Earth distance are shown. In both neutrino cases the lifetime of the mediator for best sensitivity has to be long enough to ensure that the mediator escapes the Sun, but not so long that it decays before reaching the Earth. Figure \ref{fig:4} shows the ANTARES nucleon--WIMP cross section limits for the SDM scenario with the products of DM annihilation in the Sun through mediators decaying into: di-muons (blue) and directly into neutrinos (green) for the selected mediator's lifetimes. For sufficiently long-lived, but unstable mediators, the limits imposed to these models are much more restrictive than those derived in direct detection searches for the case of spin-dependent interaction. In the case of spin-independent interactions, the direct detection search is more competitive for low and intermediate masses, but the SDM search becomes more competitive for larger masses ($> 1$ TeV). 

The limits derived here are the first experimental limits on SDM models established by a neutrino telescope. Compared to other indirect detection methods, such as those using gamma-rays, the limits derived here are in general competitive for large DM masses and favourable mediator lifetimes ($L \approx 10^{11}$ m). However, the comparison is not straightforward, since the results are usually given in terms of the $<\sigma v>$ parameter, and several astrophysical assumptions have to be made. Therefore, the different indirect searches can be considered complementary. 

\section{Conclusions}\label{concl}

An analysis of 2007-2012 ANTARES data for the search of Secluded Dark Matter annihilation from the Sun has been presented. Lately, this kind of models has increased in popularity, since SDM could help to interpret the energy spectrum of the positron/electron ratio measured recently by Pamela, Fermi-LAT and AMS-II. Assuming that DM can be annihilated through a mediator that has a long lifetime, three different situations have been considered: a) detection of dimuons that result from the mediator decay, or neutrino detection from: b) mediator that decays into a dimuon and, in turn, into neutrinos, and c) a mediator that decays directly into neutrinos. No significant excess over background has been observed for these searches, and limits to these models have been derived. This is the first time that these models are constrained with dedicated searches in neutrino telescopes. The limits are the most restrictive available for a wide range of DM masses, and mediator lifetimes, and in particular for the case of spin-dependent DM-proton interaction.

\section{Acknowledgments}

The authors acknowledge the financial support of the funding agencies:
Centre National de la Recherche Scientifique (CNRS), Commissariat \`a
l'\'ener\-gie atomique et aux \'energies alternatives (CEA),
Commission Europ\'eenne (FEDER fund and Marie Curie Program),
Institut Universitaire de France (IUF),
IdEx program and UnivEarthS Labex program at Sorbonne Paris Cit\'e
(ANR-10-LABX-0023 and ANR-11-IDEX-0005-02), R\'egion
\^Ile-de-France (DIM-ACAV), R\'egion Alsace (contrat CPER), R\'egion
Provence-Alpes-C\^ote d'Azur, D\'e\-par\-tement du Var and Ville de La
Seyne-sur-Mer, France; Bundesministerium f\"ur Bildung und Forschung
(BMBF), Germany; Istituto Nazionale di Fisica Nucleare (INFN), Italy;
Stichting voor Fundamenteel Onderzoek der Materie (FOM), Nederlandse
organisatie voor Wetenschappelijk Onderzoek (NWO), the Netherlands;
Council of the President of the Russian Federation for young
scientists and leading scientific schools supporting grants, Russia;
National Authority for Scientific Research (ANCS), Romania; 
Mi\-nis\-te\-rio de Econom\'{\i}a y Competitividad (MINECO), Prometeo 
and Grisol\'{\i}a programs of Generalitat Valenciana and MultiDark, 
Spain; Agence de  l'Oriental and CNRST, Morocco. We also acknowledge 
the technical support of Ifremer, AIM and Foselev Marine for the sea 
operation and the CC-IN2P3 for the computing facilities.

\end{document}